\documentclass{PoS}

\usepackage{cite}
\usepackage{color}
\usepackage{epsfig}
\usepackage{axodraw}
\usepackage{feynarts2}
\usepackage{wrapfig}

\newcommand{\rd}{\mathrm{d}}
\newcommand{\rL}{\mathrm{L}}
\newcommand{\rT}{\mathrm{T}}

\def\veps{\varepsilon}

\newcommand{\GeV}{\unskip\,\mathrm{GeV}}

\newcommand{\TeV}{\unskip\,\mathrm{TeV}}

\def\mathswitch#1{\relax\ifmmode#1\else$#1$\fi}
\def\mathswitchr#1{\relax\ifmmode{\mathrm{#1}}\else$\mathrm{#1}$\fi}
\def\mathswitchit#1{\relax\ifmmode{#1}\else$#1$\fi}

\newcommand{\PW}{\mathswitchr W}
\newcommand{\PZ}{\mathswitchr Z}

\newcommand{\Pg}{\mathswitchr g}
\newcommand{\PH}{\mathswitchr H}

\newcommand{\Pe}{\mathswitchr e}

\newcommand{\Pp}{\mathswitchr p}

\newcommand{\Pep}{\mathswitchr {e^+}}
\newcommand{\Pem}{\mathswitchr {e^-}}

\title{Standard Model Theory}

\ShortTitle{Standard Model Theory}

\author{\speaker{Stefan Dittmaier}%
       \\
       Universit\"at Freiburg / Physikalisches Institut\\
       E-mail: \email{stefan.dittmaier@physik.uni-freiburg.de}}

\abstract{The field of precision calculations 
for Standard Model processes at the LHC has experienced enormous
progress in recent years. This talk highlights some examples
from the physics of parton distribution functions, jets, electroweak
gauge bosons and Higgs bosons.}

\FullConference{The European Physical Society Conference on High Energy Physics\\
         5-12 July\\
         Venice, Italy}

\begin{document}

\section{Introduction}

The CERN Large Hadron Collider (LHC) was built to explore the validity of the
Standard Model (SM) of particle physics at energy scales ranging from the
electroweak (EW) scale $\sim100\GeV$ up to energies of some TeV and to search for new
phenomena and new particles in this energy domain. The discovery of a Higgs
particle at LHC Run~1 in 2012 was a first big achievement in this enterprise.
Since first studies of the properties of this Higgs particle (spin, CP parity,
couplings to the heaviest SM particles) show good agreement between measurements
and SM predictions, the SM is in better shape than ever to describe all known
particle phenomena up to very few exceptions (Dark Matter, some tension in the
measured anomalous magnetic moment of the muon, etc.). 
In view of the absence of spectacular new-physics signals in LHC data, this means
that any deviation from the SM hides in small and subtle effects. To extract
those differences from data, both experimental analyses and theoretical
predictions have to be performed with the highest possible accuracy,
i.e.\ precision can be the key to new discoveries. 
Precise theoretical predictions have to include quantum corrections, both of the
strong and EW interaction. 

This short review summarizes some recent highlights of precision
calculations for the LHC---a field that has seen enormous progress 
in the recent years.
The calculation of perturbative next-to-leading-order (NLO) QCD and EW
corrections has been successfully automated up to
particle multiplicities of roughly 4--6 (depending on the complexity of the
process) upon combining multi-purpose Monte Carlo generators or integrators with
automated one-loop matrix-element generators.
At the next-to-next-to-leading-order (NNLO) level 
QCD calculations have been completed for the most important $2\to2$ particle
scattering processes at the LHC, 
and even next-to-next-to-next-to-leading order (NNNLO) corrections
have been made possible for two Higgs-boson production channels
using specific approximations.
This progress in fixed-order calculations goes in parallel with new achievements
in the all-order calculation of leading corrections, such as
analytic QCD resummations at higher and higher levels of accuracy
and numerically working parton showers.
For example, the matching of
fixed-order calculations to QCD parton showers is meanwhile standard at NLO,
and first results exist at the NNLO level~\cite{Alioli:2013hqa};
the inclusion of photon-emission in the NLO matching has been
performed for some processes as well~\cite{Barze:2013fru}
(see also Ref.~\cite{vicini}).
Parton showers describe jet emission beyond fixed order in 
some logarithmic approximation.
For cases in which full NLO precision is desirable in higher jet multiplicities,
merging techniques are available to combine NLO calculations for
a specific event topology together with $n=0,1,2,...$ jets, 
while avoiding double-counting of jet activity.

In the following we discuss some recent advances in different directions,
including the issue of the photon density in the proton,
QCD corrections to jet physics, EW corrections to weak
gauge-boson production processes, and the global status of Higgs production
cross sections. 
Highlights from top-quark~\cite{czakon}
and flavour physics~\cite{gori}
as well as more details and examples for progress in higher-order
calculations~\cite{zanderighi,huss}
can be found elsewhere in these proceedings.

\clearpage

\section{The photon density of the proton%
\protect\footnote{See also Refs.~\cite{giuli,glazov,guffanti,sborlini}.}}

Collinear photon emission off (anti)quark partons of the proton
leads to logarithmic mass singularities in the calculation of NLO EW
corrections to partonic cross sections, just as gluon emission in
NLO QCD calculations. Analogous to the absorption of those QCD
initial-state singularities into the parton distribution functions (PDFs)
of the proton, the corresponding photonic singularities are absorbed into
the PDFs as well. This PDF redefinition naturally leads to a photon
PDF, whose dependence on its factorization scale $\mu$ is ruled by the
DGLAP evolution equations, which include the photon PDF just like
the other quark, antiquark, and gluon PDFs.

Among the usually employed PDF sets,
MRST2004qed~\cite{Martin:2004dh}
was the first that provided a photon PDF, which, however,
was completely model driven and did not include an error estimate in its original version
(a later version provides the error band shown in Fig.~\ref{fig:photonPDF} below).
About ten years later, the NNPDF group provided photon PDFs
in the NNPDF23qed~\cite{Ball:2013hta}
and NNPDF30qed~\cite{Ball:2014uwa}
PDF sets, which were derived from 
experimental data (deep-inelastic $\Pe\Pp$ scattering, Drell--Yan-like W/Z production) 
and, thus, suffered from large
errors, which could be as large as $100\%$.
Upon combining constraints from data with model assumptions, 
the photon PDF in the CT14qed set~\cite{Schmidt:2015zda}
was accurate at the level
of $10{-}20\%$.
The situation was drastically improved in 2016 with the advent
of the LUXqed photon PDF~\cite{Manohar:2016nzj},
which was derived by exploiting the trick that hadronic collisions
mediated by virtual photons only can be equivalently described by
using a photon PDF or by the parametrization of the hadronic tensor
by the structure functions $F_2$ and $F_{\mathrm{L}}$.
In this way, it is possible to derive a relation between the
photon PDF $f_\gamma(x,\mu^2)$ and the structure functions,
\begin{eqnarray}
{x f_{\gamma}(x,\mu^2)} &=& 
\frac{1}{2\pi\alpha(\mu^2)}
\int_x^1 \frac{\rd z}{z} \, \biggl\{ \,
\int^{\mu^2/(1-z)}_{x^2m_\Pp^2/(1-z)} 
\frac{\rd Q^2}{Q^2} \, \alpha(Q^2)^2 
\, \biggl[ \biggl( z p_{\gamma q}(z)+\frac{2x^2m_\Pp^2}{Q^2}\biggr)\,  
{F_2\Big(\frac{x}{z},Q^2\Big)}
\nonumber\\
&&
\hspace*{12em} 
{}-z^2 \, {F_\rL\Big(\frac{x}{z},Q^2 \Bigr)} \biggr]
-\alpha(\mu^2)^2\, z^2 \,{F_2\Big(\frac{x}{z},\mu^2\Big)} \biggr\},
\hspace{2em}
\label{eq:gammaPDF}
\end{eqnarray}
which can be numerically evaluated from data on 
$F_2(x,Q^2)$ and $F_{\mathrm{L}}(x,Q^2)$. 
In Eq.~(\ref{eq:gammaPDF}), $m_\Pp$ is the proton mass and 
$p_{\gamma q}(z)$ the $q\to\gamma q$ splitting function.
\begin{figure}[t]
\raisebox{4em}{\epsfig{file=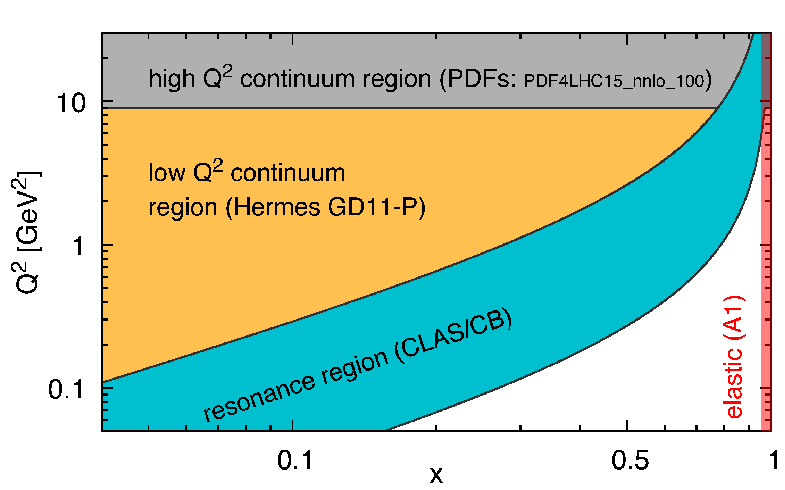,width=.5\textwidth}}
\hfill
\epsfig{file=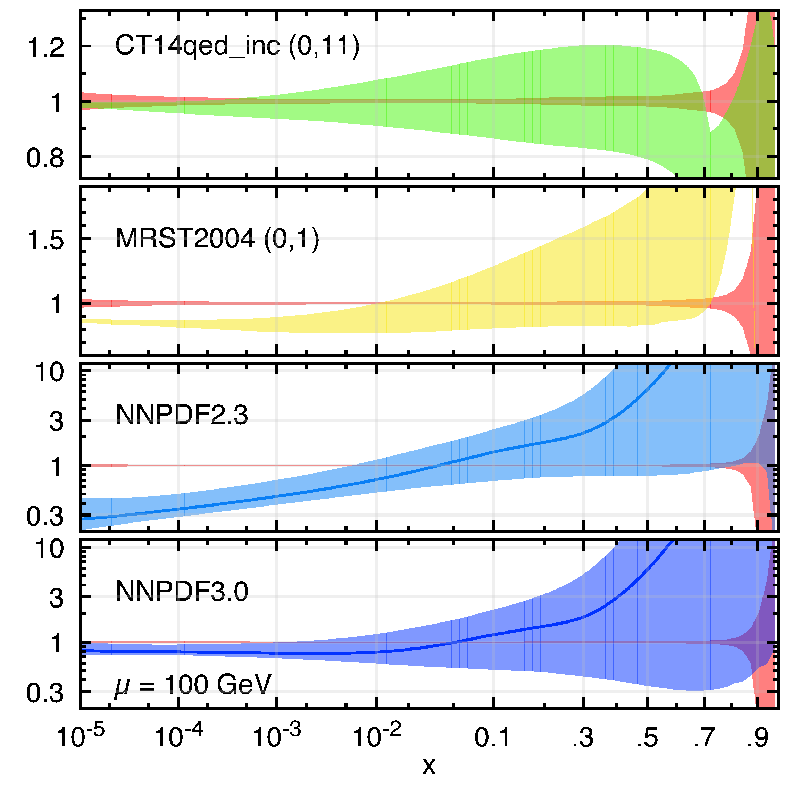,width=.5\textwidth}
\caption{Left: breakup of the $(x,Q^2)$ plane in terms of the 
$F_2(x,Q^2)$ and $F_\rL(x,Q^2)$ data used in Eq.~(\protect\ref{eq:gammaPDF}).
Right: ratio of photon PDFs from some common PDF sets (with uncertainty bands)
to the LUXqed photon PDF (uncertainty band in red).
(Taken from Ref.~\cite{Manohar:2016nzj}.)
}
\label{fig:photonPDF}
\end{figure}
The l.h.s.\ of Fig.~\ref{fig:photonPDF} illustrates the coverage
of the $(x,Q^2)$ plane by data from different experiments.
Note that the region at $x=1$ contains the contribution from
elastic scattering, where the proton does not break up in the
collision.
The r.h.s., finally, shows the comparison of the mentioned
determinations of the photon PDF, normalized to LUXqed,
with respective error bands. 
The LUXqed photon PDF is good within $1{-}2\%$ in the
typical $x$~range of LHC physics and, thus, even the best known
of all PDFs.

Partonic channels with initial-state photons exist for every
scattering reaction at the LHC, but their contribution typically
is part of the EW corrections and at the level of few percent.
Exceptions are processes where $\gamma\gamma$, $\Pg\gamma$,  or $q\gamma$ collisions
already appear in lowest order, or processes with W~bosons in the final
state. The enhancement in the latter is due to the fact that 
initial-state photons can couple to $t$-channel W~bosons, leading to
enhanced forward W~production---a mechanism that is not overwhelmed
by quark--gluon scattering. 
In the extreme case of triple-W 
production~\cite{Yong-Bai:2016sal,Dittmaier:2017bnh}, the contribution from
$q\gamma$~scattering is about $12\%$ (relative to the leading-order prediction)
at the LHC running with a
centre-of-mass (CM) energy of 
$13\TeV$.

\section{Jet production%
\protect\footnote{See also Refs.~\cite{zanderighi,trocsanyi}.}}

Investigating jet production at the LHC is not only important
as consistency check on the validity and our understanding of QCD,
it provides also important information on PDFs and another
possibility to measure the strong coupling constant $\alpha_{\mathrm{s}}$.
\begin{figure}[t]
\raisebox{3em}{\epsfig{file=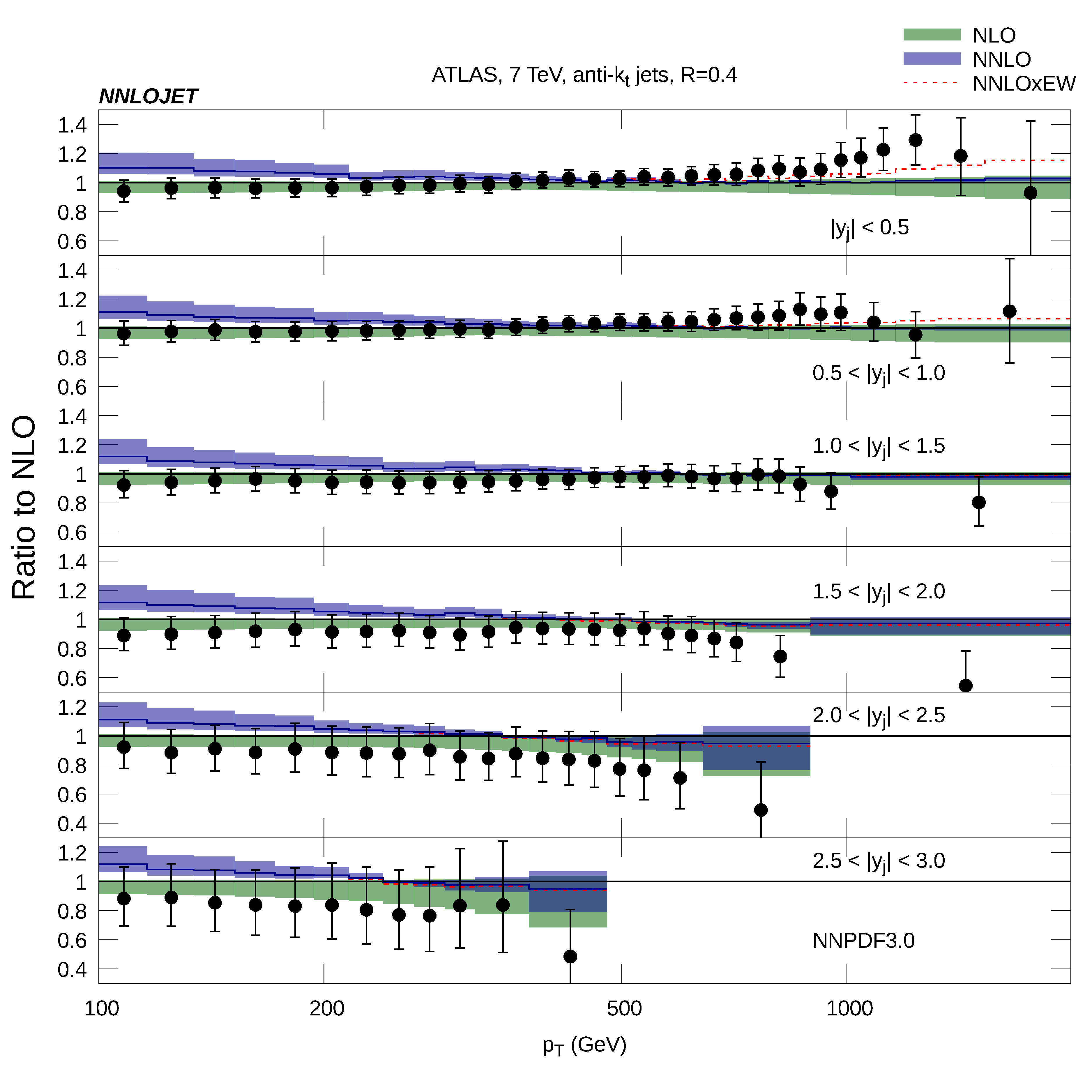,width=.5\textwidth}}
\hfill
\includegraphics[bb= 20 32 490 765,clip,width=0.45\textwidth]{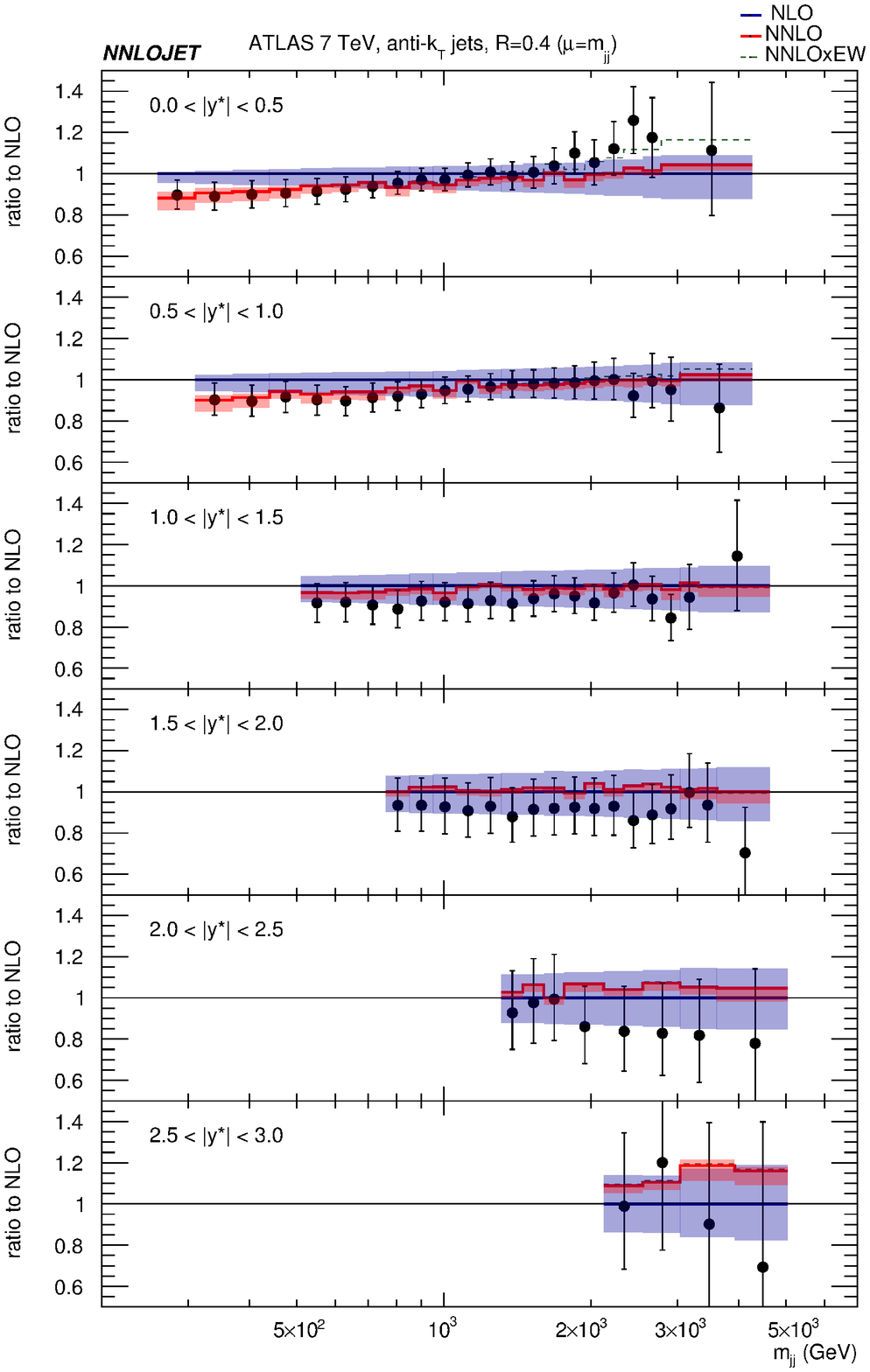}
\caption{Single-jet inclusive (left) and di-jet (right) differential
cross sections at NNLO and NLO QCD normalized to NLO QCD
in comparison to ATLAS data, with corresponding scale uncertainty
bands. The inclusion of EW corrections is
indicated by the dashed lines.
(Taken from Refs.~\cite{Currie:2016bfm,Currie:2017eqf}.)}
\label{fig:jets}
\end{figure}
Last but not least, jet production is an ubiquitous background to other
processes. 
On the theory side, it is crucial to control QCD corrections to
jet production to the highest possible level, a task that is
complicated for various reasons:
Corrections are large due to high powers of $\alpha_{\mathrm{s}}$,
jet multiplicities can be high,
a perturbatively stable definition of jets requires great care, etc..

Recently, great progress was made in predicting cross sections
for single-jet inclusive~\cite{Currie:2016bfm}
and di-jet~\cite{Currie:2017eqf} production at the
NNLO QCD level in leading-colour approximation.
Figure~\ref{fig:jets} shows some results on
the transverse-momentum ($p_\rT$) spectrum of the leading jet
in single-jet inclusive production as well as the di-jet invariant
mass ($m_{\mathrm{jj}}$) distribution in di-jet production, categorized
according to rapidity regions. 
In the transition from NLO to NNLO QCD predictions,
the scale uncertainty, which is indicated by the corresponding bands,
is reduced from typically $10{-}20\%$
to some percent at large transverse momenta or large invariant masses. 
The NNLO corrections are quite sizeable, 
and the slight tensions between NNLO prediction and data most likely are due
to the neglect or the approximative inclusion of the NNLO contributions in the 
determination of the employed PDF sets, so that a
significant impact of the new NNLO results on PDF fits can be expected.
It is also interesting to note that data start to be sensitive
to EW corrections~\cite{Dittmaier:2012kx}
as well, whose impact is indicated
by dashed lines in the plots.

\section{Electroweak gauge-boson production}

Pair production of massive EW gauge bosons is interesting
at the LHC both as signal and background process. As signal,
it bears direct information on the non-Abelian triple-gauge-boson
interactions, which are sensitive to physics beyond the SM.
As background, it is relevant to many searches for new physics
and, most notably, to analyses of Higgs bosons based on the
four-body decays $\PH\to\PW\PW/\PZ\PZ\to4\,$leptons.
Note, however, that in the latter case at least one of the two W or
Z~bosons is far off its mass shell, so that predictions based
on on-shell W- or Z-boson pairs cannot be used.

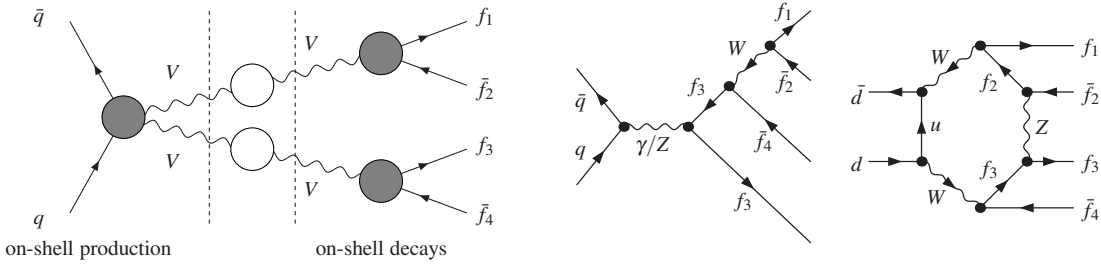
\begin{figure}[t]
{\unitlength .8pt \scriptsize \SetScale{.8}
\begin{picture}(235,120)(-25,-20)
\ArrowLine(30,50)( 5, 95)
\ArrowLine( 5, 5)(30, 50)
\Photon(30,50)(150,80){2}{11}
\Photon(30,50)(150,20){2}{11}
\ArrowLine(150,80)(190, 95)
\ArrowLine(190,65)(150,80)
\ArrowLine(190, 5)(150,20)
\ArrowLine(150,20)(190,35)
\GCirc(30,50){10}{.5}
\GCirc(90,65){10}{1}
\GCirc(90,35){10}{1}
\GCirc(150,80){10}{.5}
\GCirc(150,20){10}{.5}
\DashLine( 70,0)( 70,100){2}
\DashLine(110,0)(110,100){2}
\put(50,26){$V$}
\put(50,68){$V$}
\put(115,13){$V$}
\put(115,82){$V$}
\put(-12, 0){$q$}
\put(-12,95){$\bar q$}
\put(195, 1){$\bar f_4$}
\put(195,34){$f_3$}
\put(195,60){$\bar f_2$}
\put(195,95){$f_1$}
\put(-25,-15){on-shell production}
\put(120,-15){on-shell decays}
\end{picture}
}
\hfill
{\scriptsize
\unitlength=0.7pt%
\begin{feynartspicture}(330,150)(2,1)
\FADiagram{}
\FAProp(0.,15.)(4.,10.)(0.,){/Straight}{-1}
\FALabel(0.784783,12.8238)[tr]{$\bar q$}
\FAProp(0.,5.)(4.,10.)(0.,){/Straight}{1}
\FALabel(0.784783,7.17617)[br]{$q$}
\FAProp(20.,20.)(16.5,17.)(0.,){/Straight}{-1}
\FALabel(18.5673,19.1098)[br]{$f_1$}
\FAProp(20.,14.)(16.5,17.)(0.,){/Straight}{1}
\FALabel(18.644,14.7931)[tr]{$\bar f_2$}
\FAProp(20.,0.)(9.5,10.)(0.,){/Straight}{-1}
\FALabel(15.0576,4.29897)[tr]{$f_3$}
\FAProp(20.,7.)(13.,13.5)(0.,){/Straight}{1}
\FALabel(16.8099,9.6468)[tr]{$\bar f_4$}
\FAProp(4.,10.)(9.5,10.)(0.,){/Sine}{0}
\FALabel(6.5,9.23)[t]{$\gamma/Z$}
\FAProp(16.5,17.)(13.,13.5)(0.,){/Sine}{1}
\FALabel(14.6011,16.1387)[br]{$W$}
\FAProp(9.5,10.)(13.,13.5)(0.,){/Straight}{-1}
\FALabel(11.1011,12.6387)[br]{$f_3$}
\FAVert(4.,10.){0}
\FAVert(16.5,17.){0}
\FAVert(9.5,10.){0}
\FAVert(13.,13.5){0}
\FADiagram{}
\FAProp(1.,13.)(5.5,13.)(0.,){/Straight}{-1}
\FALabel(0.,12.77)[c]{$\bar d$}
\FAProp(1.,7.)(5.5,7.)(0.,){/Straight}{1}
\FALabel(0.,6.77)[c]{$d$}
\FAProp(18.5,17.)(10.5,17.)(0.,){/Straight}{-1}
\FALabel(20.,16.77)[c]{$f_1$}
\FAProp(18.5,13.)(14.5,13.)(0.,){/Straight}{1}
\FALabel(20.,12.77)[c]{$\bar f_2$}
\FAProp(18.5,7.)(14.5,7.)(0.,){/Straight}{-1}
\FALabel(20.,6.77)[c]{$f_3$}
\FAProp(18.5,3.)(10.5,3.)(0.,){/Straight}{1}
\FALabel(20.,2.77)[c]{$\bar f_4$}
\FAProp(5.5,13.)(5.5,7.)(0.,){/Straight}{-1}
\FALabel(6.27,10.)[l]{$u$}
\FAProp(5.5,13.)(10.5,17.)(0.,){/Sine}{-1}
\FALabel(7.97383,15.4152)[br]{$W$}
\FAProp(5.5,7.)(10.5,3.)(0.,){/Sine}{1}
\FALabel(7.68279,4.48349)[tr]{$W$}
\FAProp(10.5,17.)(14.5,13.)(0.,){/Straight}{-1}
\FALabel(12.1014,14.6014)[tr]{$f_2$}
\FAProp(14.5,13.)(14.5,7.)(0.,){/Sine}{0}
\FALabel(15.27,10.)[l]{$Z$}
\FAProp(14.5,7.)(10.5,3.)(0.,){/Straight}{-1}
\FALabel(12.1032,5.44678)[br]{$f_3$}
\FAVert(5.5,13.){0}
\FAVert(5.5,7.){0}
\FAVert(10.5,17.){0}
\FAVert(14.5,13.){0}
\FAVert(14.5,7.){0}
\FAVert(10.5,3.){0}
\end{feynartspicture}
}
\vspace*{-.5em}
\caption{Sample diagrams for four-fermion production
in $\bar qq$ annihilation.
Left: structural diagram for the factorizable contributions of a DPA;
middle: LO diagram without two intermediate W~bosons;
right: hexagon loop diagram.}
\label{fig:pp4f-diags}
\end{figure}
\begin{figure}[b]
\raisebox{1.5em}{\epsfig{file=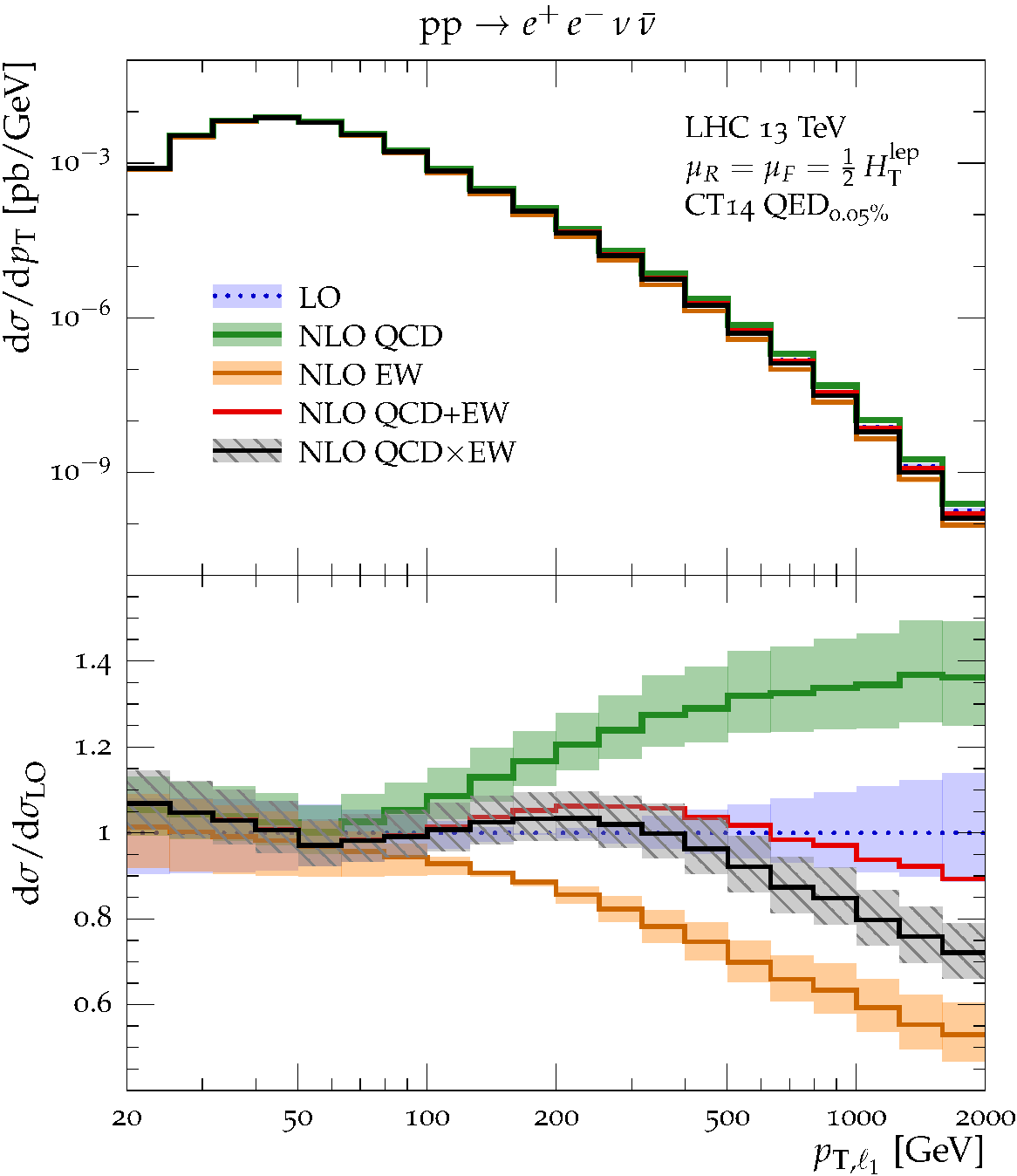,width=.5\textwidth}}
\hfill
\includegraphics[bb=75 340 370 750,width=0.45\textwidth]{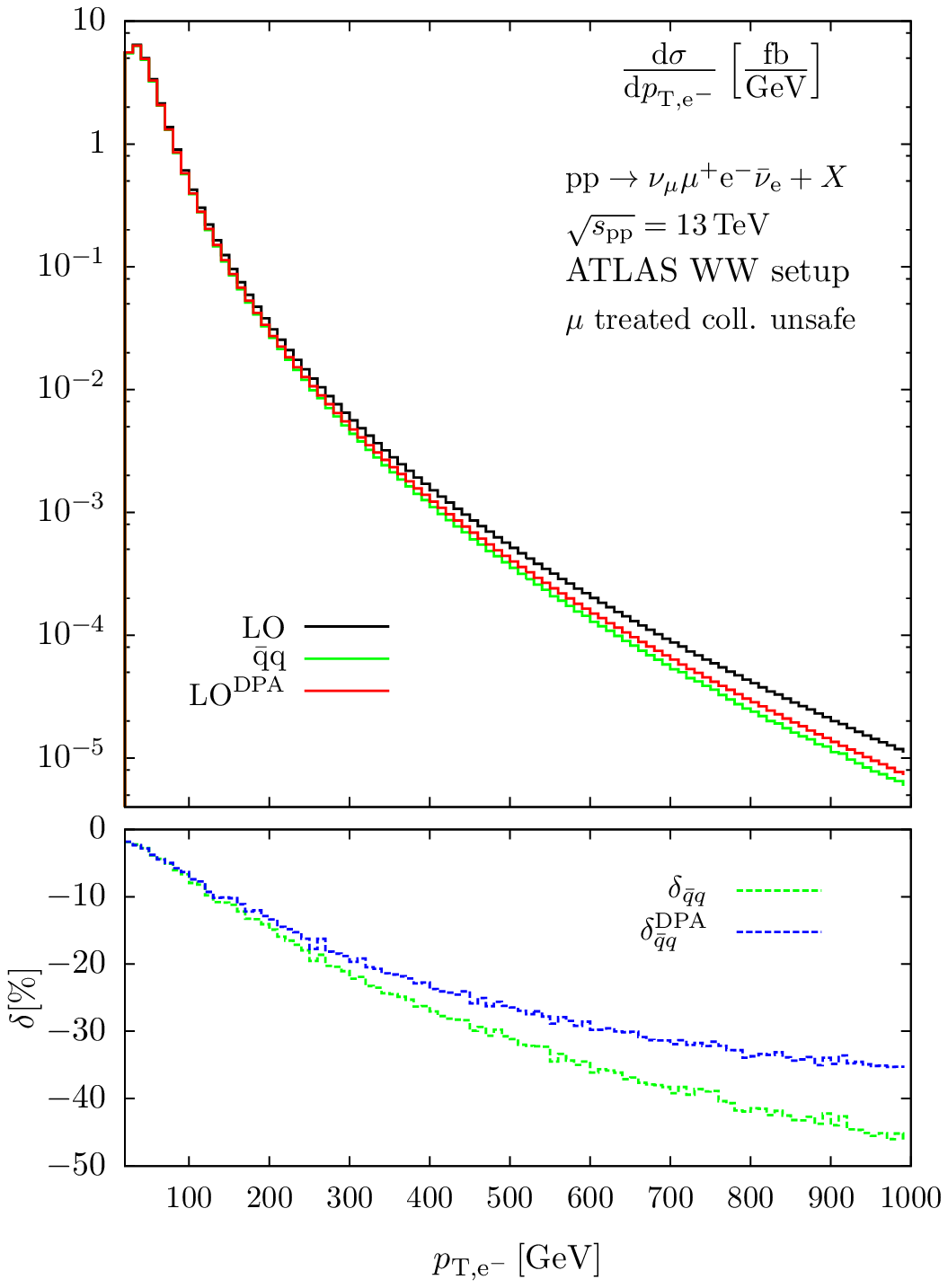}
\caption{Transverse-momentum distributions of charged leptons
in $pp\to e^+e^-\nu\bar\nu$ (left) and
in $pp\to\nu_\mu\mu^+e^-\bar\nu_e$ (right), with 
relative corrections shown in the lower panels, as described in the text.
On the r.h.s.\ NLO EW corrections to the $\bar qq$ channels obtained
from a full four-fermion calculation are compared to results from a DPA.
(Taken from Refs.~\cite{Biedermann:2016guo,Kallweit:2017khh}.)}
\label{fig:ppWW4l}
\end{figure}

In previous years, the theoretical descriptions of these processes
made major leaps: 
QCD predictions were pushed to 
NNLO~\cite{Cascioli:2014yka}
(+ $gg$ channels to NLO~\cite{Caola:2015psa}),
and EW NLO corrections, which were only known for 
on-shell W/Z~bosons~\cite{Bierweiler:2012kw} or
in the form of resonance expansions~\cite{Billoni:2013aba}
before,
were generalized to full off-shell calculations for 
four-lepton production~\cite{Biedermann:2016yvs,Biedermann:2016guo,%
Kallweit:2017khh,Biedermann:2017oae}
using the complex-mass scheme~\cite{Denner:2005fg}
for a gauge-invariant treatment
of the resonances.
Figure~\ref{fig:pp4f-diags} illustrates on its l.h.s.\
the structure of loop diagrams
of the so-called {\it factorizable} corrections
which furnish the dominating contributions in an expansion
of amplitudes about the resonance poles. 
The leading term
of such an expansion is known as {\it double-pole approximation} (DPA).
The middle and left diagrams in Fig.~\ref{fig:pp4f-diags} show tree-level
and one-loop diagrams that would be neglected in a leading-order (LO) or NLO DPA,
respectively.
The l.h.s.\ of Fig.~\ref{fig:ppWW4l} shows the NLO QCD and EW corrections
to the transverse-momentum spectrum of a charged lepton 
in the process $\Pp\Pp\to\Pep\Pem\nu\bar\nu$~\cite{Kallweit:2017khh}, 
which is dominated by
W-pair production. 
The EW corrections show the known Sudakov enhancement
to several $10\%$ in the TeV range, a regime explored by the LHC
deeper and deeper in the next years.
The r.h.s.\ of the figure compares results of the full
$2\to4$ off-shell calculation~\cite{Biedermann:2016guo} of the EW corrections 
to $\Pp\Pp\to\nu_\mu\mu^+\Pem\bar\nu_\Pe$ with a corresponding 
DPA~\cite{Billoni:2013aba}. While the DPA represents a very good
approximation whenever two W~resonances dominate
(integrated cross sections, rapidity distributions, 
momentum spectra at smaller energies, etc.),
it fails for transverse lepton momenta at high 
\begin{wrapfigure}{r}{0.5\textwidth}
\includegraphics[width=0.5\textwidth]{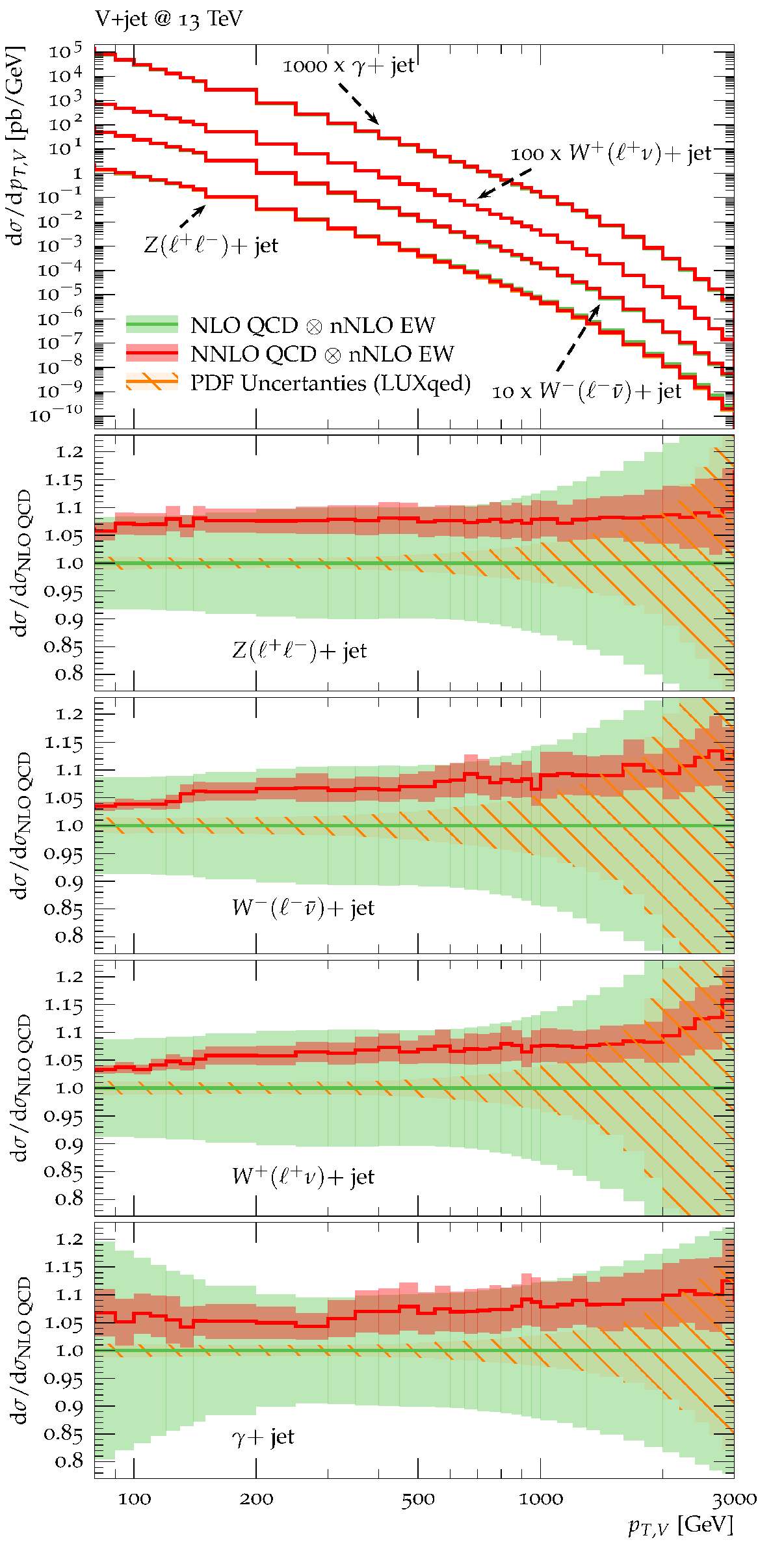}
\caption{$V+$jet spectra predicted at (N)NLO QCD $\otimes$ nNLO EW,
where the lower panels show the relative impact of 
(N)NLO QCD $\otimes$ nNLO EW over NLO QCD $\otimes$ nNLO EW,
together with the corresponding uncertainty bands
(nNLO EW is the sum of NLO EW and
leading NNLO EW Sudakov logarithms).
PDF uncertainties are included as hashed orange bands.
(Taken from Ref.~\cite{Lindert:2017olm}.)}
\vspace*{-2em}
\label{fig:V+jet}
\end{wrapfigure}
energies,
where the DPA misses corrections to singly-resonant diagrams 
as shown in the middle of Fig.~\ref{fig:pp4f-diags}, which
can be viewed as W~bremsstrahlung to Drell--Yan-like lepton pair production.

Figure~\ref{fig:ppWW4l} (left) also illustrates the issue of
combining QCD and EW corrections by comparing the two extreme variants
of simply adding (QCD$+$EW) or factorizing (QCD$\times$EW) 
QCD and EW corrections. 
If both types of corrections become large
the differences between the two possibilities can get large,
but the difference itself is an overly conservative estimate of
the uncertainty of the combination, 
because leading corrections are known to factorize
to a large extent.
This and other issues in the determination of a realistic estimate
of theory and parametric uncertainties in predictions 
were
analyzed in Ref.~\cite{Lindert:2017olm}, where 
EW vector-boson+jet production was considered as background to
Dark Matter searches at high transverse vector-boson momenta $p_{\rT,V}$
at the LHC.
Schematically, Monte Carlo (MC) and theory (TH) uncertainties are
introduced in Monte Carlo predictions via reweighting,
\looseness-1
\begin{eqnarray}
\lefteqn{\frac{\rd}{\rd x}\,\frac{\rd}{\rd\vec y}\,
\sigma(\vec\veps_{\mathrm{MC}},{\vec\veps_{\mathrm{TH}}})} &&
\\
&=&
\frac{\rd}{\rd x}\,\frac{\rd}{\rd\vec y}\,
\sigma_{\mathrm{MC}}(\vec\veps_{\mathrm{MC}})
\,  \times \,
\left( \frac{ \frac{\rd}{\rd x}\, 
{\sigma_{\mathrm{TH}}}({\vec\veps_{\mathrm{TH}}}) }
             { \frac{\rd}{\rd x}\, \sigma_{\mathrm{MC}}(\vec\veps_{\mathrm{MC}}) } 
\right), 
\nonumber
\end{eqnarray}
where fully differential MC cross sections $\sigma_{\mathrm{MC}}$
are reweighted by the ratio to a state-of-the-art TH prediction
$\sigma_{\mathrm{TH}}$ which is only differential in one
(but distinctive) kinematical variable $x$, which was taken to be
$x=p_{\rT,V}$ ($\vec y$ denotes the remaining phase-space variables).
The nuisance parameters $\vec\veps_{\mathrm{MC/TH}}$ control
the impact of scale uncertainties, missing higher-order effects,
the combination of QCD and EW corrections,
PDF uncertainties, and 
correlations between different processes ($\PW/\PZ/\gamma+{}$jet)
and phase-space regions.
Figure~\ref{fig:V+jet} illustrates the resulting uncertainties,
revealing surprisingly good precision even for $p_{\rT,V}$ in the TeV range.
Perturbative cross-section uncertainties (combined quadratically) are about $5\%$
and PDF uncertainties (correlated among processes) about
$5{-}10\%$, leading to a W/Z cross-section ratio uncertainty (not shown)
of about $1{-}2\%$.
The detailed results from this study are, of course, specific to
$\PW/\PZ/\gamma+$jet production, but the methodology can be transferred
to other processes.

As a last example from EW vector-boson physics, we consider the
scattering of massive EW vector-bosons (see also Ref.~\cite{maina}), 
such as $\PW\PW\to\PW\PW$,
which is not only sensitive to triple and quartic gauge-boson
self-interactions, but also to the mechanism of EW symmetry breaking
via off-shell Higgs boson exchange in spontaneously broken gauge theories
(or any other related effect in other theories).
These processes are part of $VV+2$jet production processes 
at the LHC,
which has reported first successful analyses for like-sign
W-boson pairs.
The existence of 4~leptons and 2~jets in the final state
renders theoretical predictions with higher-order corrections
to such processes extremely demanding. 
\begin{figure}[t]
{\setlength{\unitlength}{.8pt}
\begin{picture}(120,110)(-5,10)
\SetScale{0.8}
\ArrowLine( 10, 15)( 30, 15)
\ArrowLine( 30, 15)(110, 15)
\ArrowLine(110,115)( 30,115)
\ArrowLine( 30,115)( 10,115)
\ArrowLine(110, 30)( 95, 45)
\ArrowLine( 95, 45)(110, 55)
\ArrowLine(110, 75)( 95, 85)
\ArrowLine( 95, 85)(110,100)
\Photon(30, 15)(60, 65){2}{7}
\Photon(30,115)(60, 65){2}{7}
\Photon(95, 45)(60, 65){2}{6}
\Photon(95, 85)(60, 65){2}{6}
\Vertex(30, 15){2.0}
\Vertex(30,115){2.0}
\Vertex(95, 85){2.0}
\Vertex(95, 45){2.0}
\Vertex(60, 65){2.0}
\put(27,81){$\PW$}
\put(27,41){$\PW$}
\put(70,84){$\PW$}
\put(70,38){$\PW$}
\end{picture}
\begin{picture}(120,110)(-5,10)
\SetScale{0.8}
\ArrowLine( 10, 15)( 30, 15)
\ArrowLine( 30, 15)( 70, 15)
\ArrowLine( 70, 15)(110, 15)
\ArrowLine( 70,115)( 30,115)
\ArrowLine(110,115)( 70,115)
\ArrowLine( 30,115)( 10,115)
\ArrowLine(110, 30)( 95, 45)
\ArrowLine( 95, 45)(110, 55)
\ArrowLine(110, 75)( 95, 85)
\ArrowLine( 95, 85)(110,100)
\Gluon(30, 15)(30,115){2}{14}
\Vertex(30, 15){2.0}
\Vertex(30,115){2.0}
\Photon(95, 45)(70, 15){2}{6}
\Photon(95, 85)(70,115){2}{6}
\Vertex(70, 15){2.0}
\Vertex(70,115){2.0}
\Vertex(95, 85){2.0}
\Vertex(95, 45){2.0}
\put(15,61){{$\Pg$}}
\put(65,88){$\PW$}
\put(65,31){$\PW$}
\end{picture}
\begin{picture}(140,110)(-25,10)
\SetScale{0.8}
\ArrowLine(-10, 15)( 10, 15)
\ArrowLine( 10, 15)( 30, 15)
\ArrowLine( 30, 15)(110, 15)
\ArrowLine(110,115)( 30,115)
\ArrowLine( 30,115)( 10,115)
\ArrowLine( 10,115)(-10,115)
\ArrowLine(110, 30)( 95, 45)
\ArrowLine( 95, 45)(110, 55)
\ArrowLine(110, 75)( 95, 85)
\ArrowLine( 95, 85)(110,100)
\Vertex(10,115){2.0}
\Vertex(10, 15){2.0}
\Photon(10, 15)(10,115){2}{14}
\Photon(30, 15)(60, 65){2}{7}
\Photon(30,115)(60, 65){2}{7}
\Photon(95, 45)(60, 65){2}{6}
\Photon(95, 85)(60, 65){2}{6}
\Vertex(30, 15){2.0}
\Vertex(30,115){2.0}
\Vertex(95, 85){2.0}
\Vertex(95, 45){2.0}
\Vertex(60, 65){2.0}
\put(25,86){$\PW$}
\put(25,33){$\PW$}
\put(70,84){$\PW$}
\put(70,37){$\PW$}
\put(17,68){$\gamma/\PZ$}
\put(19,52){$/\PW$}
\end{picture}
\begin{picture}(140,110)(-25,10)
\SetScale{0.8}
\ArrowLine(-10, 15)( 10, 15)
\ArrowLine( 10, 15)( 40, 15)
\ArrowLine( 40, 15)( 70, 15)
\ArrowLine( 70, 15)(110, 15)
\ArrowLine( 70,115)( 40,115)
\ArrowLine(110,115)( 70,115)
\ArrowLine( 40,115)( 10,115)
\ArrowLine( 10,115)(-10,115)
\ArrowLine(110, 30)( 95, 45)
\ArrowLine( 95, 45)(110, 55)
\ArrowLine(110, 75)( 95, 85)
\ArrowLine( 95, 85)(110,100)
\Photon(40, 15)(40,115){2}{14}
\Vertex(40, 15){2.0}
\Vertex(40,115){2.0}
\Photon(10, 15)(10,115){2}{14}
\Vertex(10, 15){2.0}
\Vertex(10,115){2.0}
\Photon(95, 45)(70, 15){2}{6}
\Photon(95, 85)(70,115){2}{6}
\Vertex(70, 15){2.0}
\Vertex(70,115){2.0}
\Vertex(95, 85){2.0}
\Vertex(95, 45){2.0}
\put(48,60){$\gamma/\PZ/\PW$}
\put(65,90){$\PW$}
\put(65,28){$\PW$}
\end{picture}
}
\caption{Sample LO and NLO diagrams for vector-boson scattering at the LHC.}
\label{fig:VBS-diags}
\end{figure}
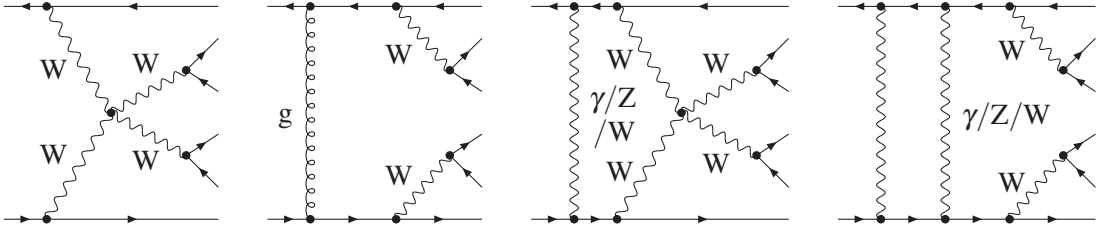
\begin{figure}[b]
\hspace*{-3em}
\epsfig{file=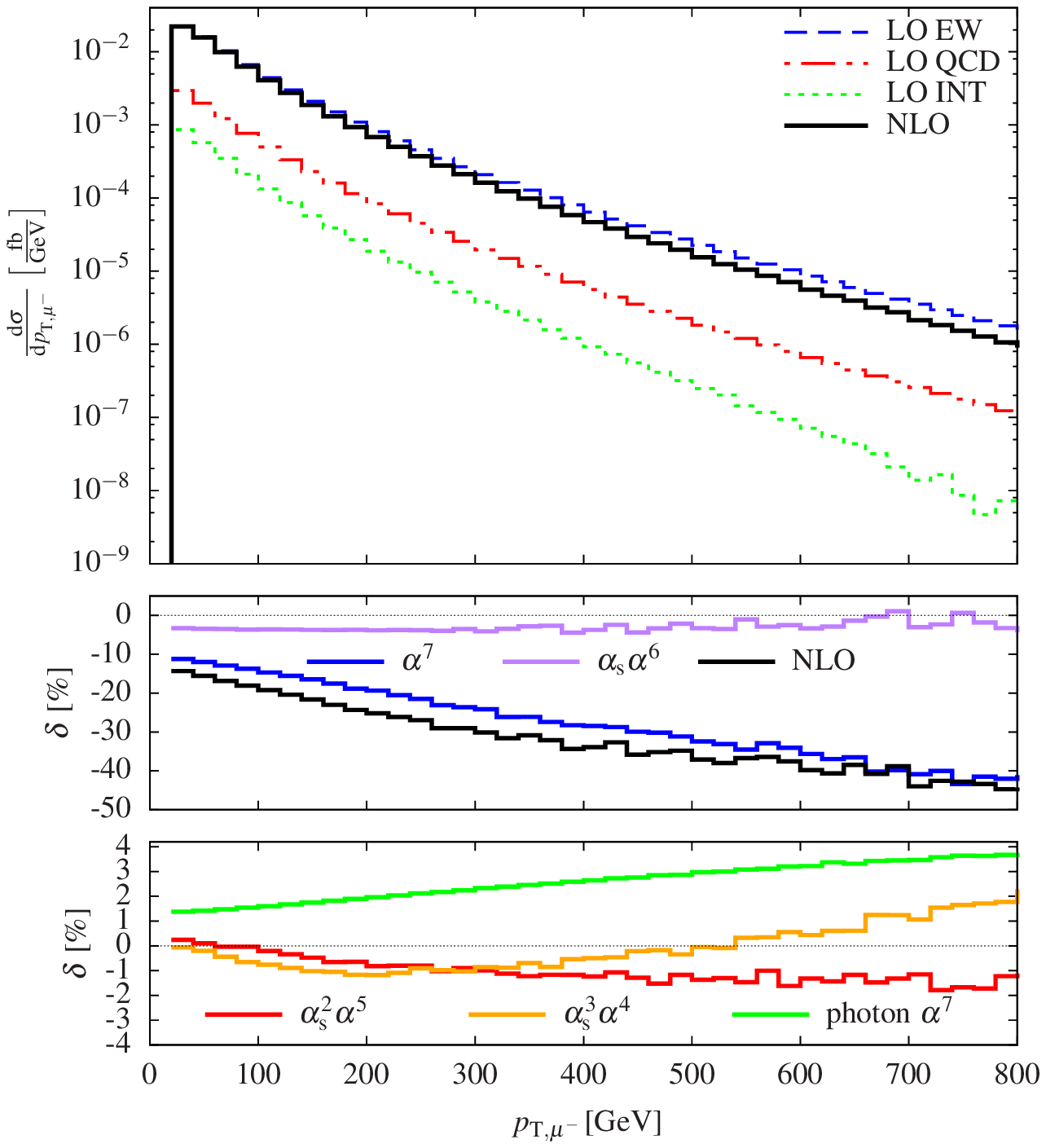,width=.55\textwidth}
\hfill
\hspace*{-2.5em}
\epsfig{file=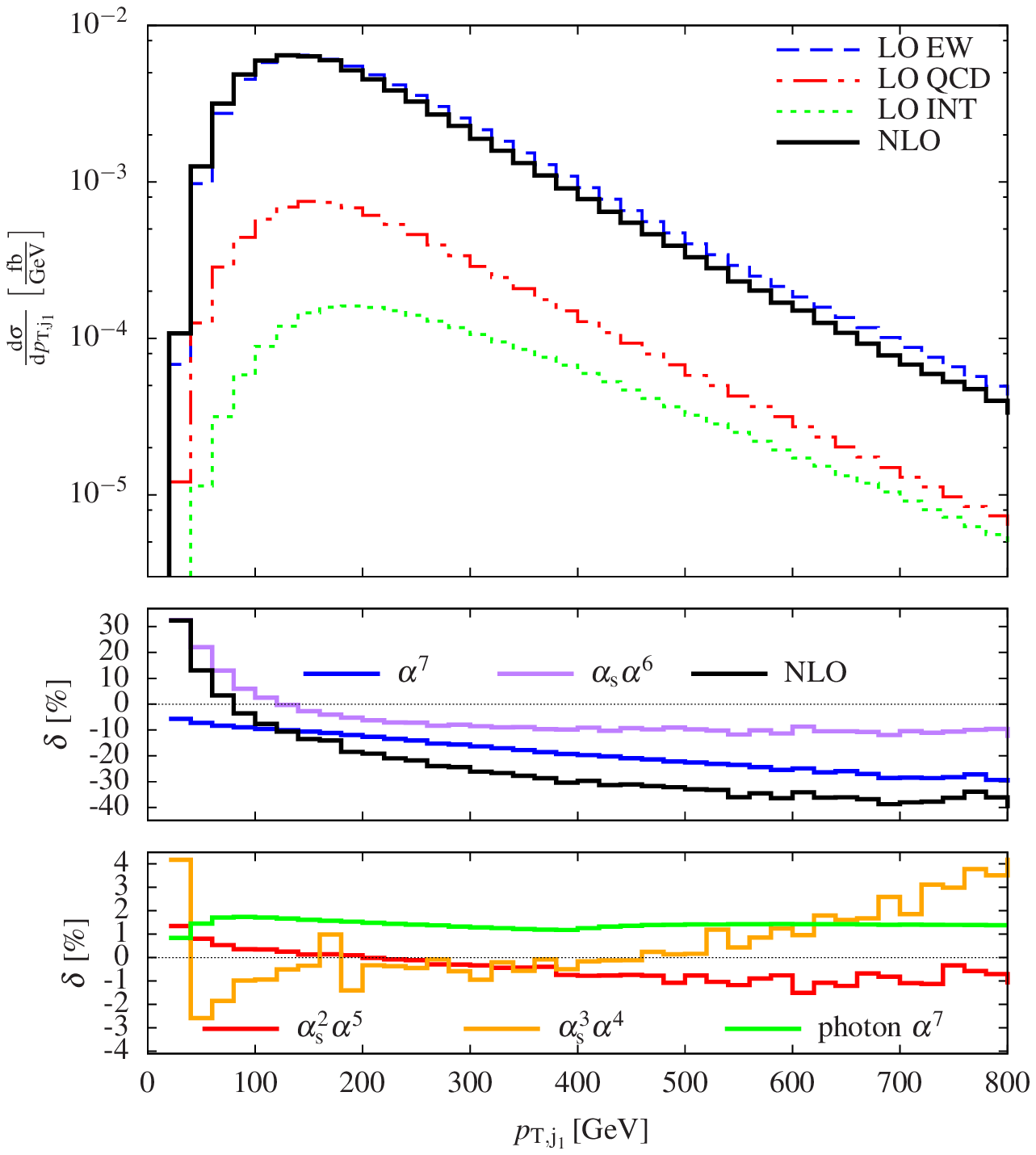,width=.55\textwidth}
\caption{Transverse-momentum distributions of the muon (left) and
the leading jet (right) and corresponding relative corrections (lower panels)
for the process $pp\to\mu^+\nu_\mu e^+\nu_e\mathrm{jj}+X$ at the LHC
with CM energy 13~TeV.
(Taken from Ref.~\cite{Biedermann:2017bss}.)}
\label{fig:VBSresults}
\end{figure}
Recently first results from an NLO calculation for the full
$2\to6$ particle process involving like-sign W~pairs were presented
in Refs.~\cite{Biedermann:2016yds,Biedermann:2017bss},
based on the one-loop matrix-element generator 
{\sc Recola}~\cite{Actis:2012qn} and the numerical one-loop integral
library {\sc Collier}~\cite{Denner:2016kdg}.
The diagrams in
Fig.~\ref{fig:VBS-diags} show that already at LO different
perturbative orders $\propto\alpha_{\mathrm{s}}^m\alpha^n$ 
contribute to the cross section, leading to four different orders
at NLO. 
The pure QCD corrections to the two LO channels indicated in 
Fig.~\ref{fig:VBS-diags} were already calculated in Refs.~\cite{Jager:2009xx}e

In Fig.~\ref{fig:VBSresults} the impact of those orders is shown
separately for the transverse-momentum spectra of a charged lepton
and the leading outgoing jet for typical selection cuts for vector-boson
scattering.
As expected, the contributions 
$\propto\alpha_{\mathrm{s}}\alpha^6$ or $\alpha^7$
with the highest powers in the EW couplings
deliver the largest corrections, with up to some $10\%$ when the TeV range
is approached, rendering those contributions important in future data
analyses. On the other hand, photon-induced channels
(shown as green lines) contribute to the signal only a few percent.

\section{Higgs-boson production}

\begin{figure}[b]
\raisebox{-8em}{\epsfig{file=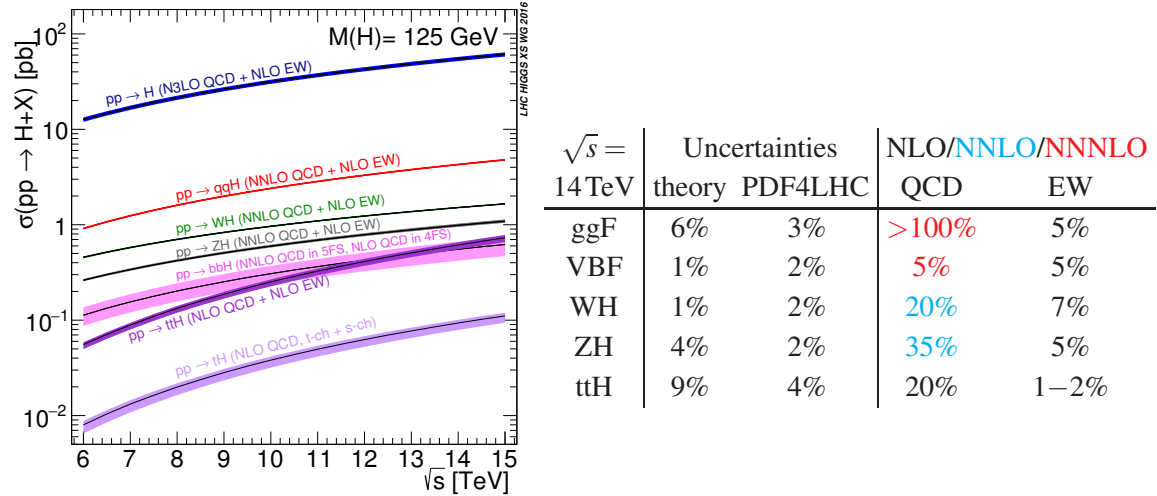,width=.47\textwidth}}
\hspace*{-1em}
{
\renewcommand{\tabcolsep}{0.3em}
\begin{tabular}{c|cc|cc}
$\sqrt{s}=$
& \multicolumn{2}{c|}{Uncertainties} &
\multicolumn{2}{c}{NLO/\Cyan{NNLO}/\Red{NNNLO}}
\\[-.0em]
$14\TeV$
& theory & $\!$PDF4LHC & QCD & {EW}
\\
\hline
ggF & {$6\%$} & $3\%$ & \Red{${>}100\%$} & {$5\%$}
\\
VBF & $1\%$ & $2\%$ & \Red{$5\%$} & {$5\%$}
\\
WH  & $1\%$ & $2\%$ & \Cyan{$20\%$} & {$7\%$}
\\
ZH  & $4\%$ & $2\%$ & \Cyan{$35\%$} & {$5\%$}
\\
ttH & $9\%$ & $4\%$ & $20\%$ & {$1{-}2\%$}
\end{tabular}
}
\vspace*{-1em}
\caption{Higgs production
cross sections at the LHC and corresponding uncertainty bands (left),
as given by the LHCHXSWG~\cite{deFlorian:2016spz},
and estimates of theory and PDF uncertainties as well as the
typical size of QCD and EW corrections for the most important 
channels.}
\label{fig:higgs}
\end{figure}

Since many years the LHC Higgs Cross Section Working Group provides
state-of-the-art predictions for the various Higgs-boson production
and decay channels, as well as details of many other phenomenological
aspects and strategies in Higgs physics.
As an important example taken from the recent CERN
Yellow Report~\cite{deFlorian:2016spz},
the l.h.s.\ of Figure~\ref{fig:higgs}
shows an overview of the cross sections of various Higgs-boson production 
channels in the SM as function of the CM energy $\sqrt{s}$ of
a pp~collider, together with the bands reflecting the
combined theoretical and PDF uncertainties.
The table on the r.h.s.\ compiles the orders of magnitude of the
uncertainties and the impact of QCD and EW corrections for the most
important channels. Impressively, the uncertainties in the QCD-driven
channels (ggF, ttH) are below $10\%$ (see also Ref.~\cite{stebel})
and in the EW-driven channels
(VBF, WH, ZH) below $5\%$. 
To reach this high level of accuracy, in most cases QCD corrections
beyond NLO and EW corrections at NLO are required.
For details, we have to refer to Ref.~\cite{deFlorian:2016spz} 
and references therein.
As highlights, we just mention the level of NNNLO in QCD 
achieved for gluon--gluon fusion (ggF)~\cite{Anastasiou:2015ema}
and vector-boson fusion (VBF)~\cite{Dreyer:2016oyx}
(although the latter is not yet included on the l.h.s.\
of Fig.~\ref{fig:higgs}) in appropriate approximations.


\clearpage

\begin{thebibliography}{99}

\bibitem{Alioli:2013hqa}
  S.~Alioli {\it et al.},
  JHEP {\bf 1406} (2014) 089
  [arXiv:1311.0286 [hep-ph]] and
%
  Phys.\ Rev.\ D {\bf 92} (2015) no.9,  094020
  [arXiv:1508.01475 [hep-ph]]; \\
%
  S.~H\"oche {\it et al.},
  Phys.\ Rev.\ D {\bf 91} (2015) no.7,  074015
  [arXiv:1405.3607 [hep-ph]] and
%
  Phys.\ Rev.\ D {\bf 90} (2014) no.5,  054011
  [arXiv:1407.3773 [hep-ph]].

\bibitem{Barze:2013fru}
  L.~Barze {\it et al.},
  Eur.\ Phys.\ J.\ C {\bf 73} (2013) no.6,  2474
  [arXiv:1302.4606 [hep-ph]]; \\
%
  A.~M\"uck and L.~Oymanns,
  JHEP {\bf 1705} (2017) 090
  [arXiv:1612.04292 [hep-ph]]; \\
%
  C.~M.~Carloni Calame {\it et al.},
  arXiv:1612.02841 [hep-ph].

\bibitem{vicini} A.~Vicini, these proceedings.
\bibitem{czakon} M.~Czakon, these proceedings.
\bibitem{gori} S.~Gori, these proceedings.
\bibitem{zanderighi} G.~Zanderighi, these proceedings.
\bibitem{huss} A.~Huss, these proceedings.
\bibitem{giuli} F.~Giuli, these proceedings.
\bibitem{glazov} A.~Glazov, these proceedings.
\bibitem{guffanti} A.~Guffanti, these proceedings.
\bibitem{sborlini} G.~Sborlini, these proceedings.

\bibitem{Martin:2004dh}
  A.~D.~Martin {\it et al.},
  Eur.\ Phys.\ J.\ C {\bf 39} (2005) 155
  [hep-ph/0411040].

\bibitem{Ball:2013hta}
  R.~D.~Ball {\it et al.} [NNPDF Collaboration],
  Nucl.\ Phys.\ B {\bf 877} (2013) 290
  [arXiv:1308.0598 [hep-ph]].

\bibitem{Ball:2014uwa}
  R.~D.~Ball {\it et al.} [NNPDF Collaboration],
  JHEP {\bf 1504} (2015) 040
  [arXiv:1410.8849 [hep-ph]].

\bibitem{Schmidt:2015zda}
  C.~Schmidt {\it et al.},
  Phys.\ Rev.\ D {\bf 93} (2016) no.11,  114015
  [arXiv:1509.02905 [hep-ph]].

\bibitem{Manohar:2016nzj}
  A.~Manohar {et al.},
  Phys.\ Rev.\ Lett.\  {\bf 117} (2016) no.24,  242002
  [arXiv:1607.04266 [hep-ph]] and
%
  arXiv:1708.01256 [hep-ph].

\bibitem{Yong-Bai:2016sal}
  Y.~B.~Shen {et al.},
  Phys.\ Rev.\ D {\bf 95} (2017) no.7,  073005
  [arXiv:1605.00554 [hep-ph]].

\bibitem{Dittmaier:2017bnh}
  S.~Dittmaier {et al.},
  JHEP {\bf 1709} (2017) 034
  [arXiv:1705.03722 [hep-ph]].

\bibitem{trocsanyi} Z.~Trocsanyi, these proceedings.

\bibitem{Currie:2016bfm}
  J.~Currie {et al.},
  Phys.\ Rev.\ Lett.\  {\bf 118} (2017) no.7,  072002
  [arXiv:1611.01460 [hep-ph]].

\bibitem{Currie:2017eqf}
  J.~Currie {et al.},
  arXiv:1705.10271 [hep-ph].

\bibitem{Dittmaier:2012kx}
  S.~Dittmaier {et al.},
  JHEP {\bf 1211} (2012) 095
  [arXiv:1210.0438 [hep-ph]]; \\
%
  J.~M.~Campbell {et al.},
  Phys.\ Rev.\ D {\bf 94} (2016) no.9,  093009
  [arXiv:1608.03356 [hep-ph]]; \\
%
  R.~Frederix {et al.},
  JHEP {\bf 1704} (2017) 076
  [arXiv:1612.06548 [hep-ph]].

\bibitem{Cascioli:2014yka}
  F.~Cascioli {\it et al.},
  Phys.\ Lett.\ B {\bf 735} (2014) 311
  [arXiv:1405.2219 [hep-ph]]; \\
%
  T.~Gehrmann {\it et al.},
  Phys.\ Rev.\ Lett.\  {\bf 113} (2014) 21,  212001
  [arXiv:1408.5243 [hep-ph]];\\
%
  M.~Grazzini {\it et al.},
  Phys.\ Lett.\ B {\bf 750} (2015) 407
  [arXiv:1507.06257 [hep-ph]]; 
%
  JHEP {\bf 1608} (2016) 140
  [arXiv:1605.02716 [hep-ph]] and
%
  JHEP {\bf 1705} (2017) 139
  [arXiv:1703.09065 [hep-ph]].

\bibitem{Caola:2015psa}
  F.~Caola {et al.},
  Phys.\ Rev.\ D {\bf 92} (2015) no.9,  094028
  [arXiv:1509.06734 [hep-ph]];
%
  Phys.\ Lett.\ B {\bf 754} (2016) 275
  [arXiv:1511.08617 [hep-ph]] and
%
  JHEP {\bf 1607} (2016) 087
  [arXiv:1605.04610 [hep-ph]].

\bibitem{Bierweiler:2012kw}
  A.~Bierweiler {et al.},
  JHEP {\bf 1211} (2012) 093
  [arXiv:1208.3147 [hep-ph]].
%
  JHEP {\bf 1312} (2013) 071
  [arXiv:1305.5402 [hep-ph]]; \\
%
  J.~Baglio {et al.},
  Phys.\ Rev.\ D {\bf 88} (2013) 113005
   Erratum: [Phys.\ Rev.\ D {\bf 94} (2016) no.9,  099902]
  [arXiv:1307.4331 [hep-ph]].


\bibitem{Billoni:2013aba}
  M.~Billoni {et al.},
  JHEP {\bf 1312} (2013) 043
  [arXiv:1310.1564 [hep-ph]].

\bibitem{Biedermann:2016yvs}
  B.~Biedermann {et al.},
  Phys.\ Rev.\ Lett.\  {\bf 116} (2016) no.16,  161803
  [arXiv:1601.07787 [hep-ph]] and
%
  JHEP {\bf 1701} (2017) 033
  [arXiv:1611.05338 [hep-ph]].

\bibitem{Biedermann:2016guo}
  B.~Biedermann {et al.},
  JHEP {\bf 1606} (2016) 065
  [arXiv:1605.03419 [hep-ph]].

\bibitem{Kallweit:2017khh}
  S.~Kallweit {et al.},
  arXiv:1705.00598 [hep-ph].

\bibitem{Biedermann:2017oae}
  B.~Biedermann {et al.},
  arXiv:1708.06938 [hep-ph].

\bibitem{Denner:2005fg}
  A.~Denner {et al.},
  Nucl.\ Phys.\ B {\bf 724} (2005) 247
   E: [Nucl.\ Phys.\ B {\bf 854} (2012) 504]
  [hep-ph/0505042].

\bibitem{Lindert:2017olm}
  J.~M.~Lindert {\it et al.},
  arXiv:1705.04664 [hep-ph].

\bibitem{maina} E.~Maina, these proceedings.

\bibitem{Biedermann:2016yds}
  B.~Biedermann {et al.},
  Phys.\ Rev.\ Lett.\  {\bf 118} (2017) no.26,  261801
  [arXiv:1611.02951 [hep-ph]].

\bibitem{Biedermann:2017bss}
  B.~Biedermann {et al.},
  arXiv:1708.00268 [hep-ph].

\bibitem{Actis:2012qn}
  S.~Actis {\it et al.},
  JHEP {\bf 1304} (2013) 037
  [arXiv:1211.6316 [hep-ph]] and
%
  Comput.\ Phys.\ Commun.\  {\bf 214} (2017) 140
  [arXiv:1605.01090 [hep-ph]].

\bibitem{Denner:2016kdg}
  A.~Denner {\it et al.},
  Comput.\ Phys.\ Commun.\  {\bf 212} (2017) 220
  [arXiv:1604.06792 [hep-ph]].

\bibitem{Jager:2009xx}
B.~J\"ager {\it et al.},
  Phys.\ Rev.\  {\bf D80 } (2009)  034022.
  [arXiv:0907.0580 [hep-ph]];\\
%
  T.~Melia {\it et al.},
  JHEP {\bf 1012} (2010) 053
  [arXiv:1007.5313 [hep-ph]] and
%
  Eur.\ Phys.\ J.\  {\bf C71 } (2011)  1670
  [arXiv:1102.4846 [hep-ph]];\\
%
%
  B.~J\"ager and G.~Zanderighi,
  JHEP {\bf 1111} (2011) 055
  [arXiv:1108.0864 [hep-ph]];\\
%
  A.~Denner {\it et al.},
  Phys.\ Rev.\ D {\bf 86} (2012) 114014
  [arXiv:1209.2389 [hep-ph]].

\bibitem{deFlorian:2016spz}
  D.~de Florian {\it et al.} [LHC Higgs Cross Section Working Group],
  arXiv:1610.07922 [hep-ph].

\bibitem{stebel} T.~Stebel, these proceedings.

\bibitem{Anastasiou:2015ema}
  C.~Anastasiou {\it et al.},
  Phys.\ Rev.\ Lett.\  {\bf 114} (2015) 212001
  [arXiv:1503.06056 [hep-ph]] and
%
  JHEP {\bf 1605} (2016) 058
  [arXiv:1602.00695 [hep-ph]].

\bibitem{Dreyer:2016oyx}
  F.~A.~Dreyer and A.~Karlberg,
  Phys.\ Rev.\ Lett.\  {\bf 117} (2016) no.7,  072001
  [arXiv:1606.00840 [hep-ph]].

\end{thebibliography}
\end{document}